\documentclass[pre,superscriptaddress,twocolumn,aps,10pt]{revtex4-1}
\pdfoutput=1
\usepackage[colorlinks=true,urlcolor=blue]{hyperref}
\usepackage{amsfonts}
\usepackage{graphicx}
\usepackage{mathrsfs}
\usepackage{amsmath}
\usepackage{xcolor}
\usepackage{bm}
\usepackage{upgreek}

\usepackage{color}
\definecolor{red}{rgb}{0.75,0,0}
\definecolor{blue}{rgb}{0,0,0.75}
\definecolor{green}{rgb}{0,0.5,0}

\newcommand{\dd}{\mathrm{d}}

\begin{document}
\title{Topotaxis of active Brownian particles}

\author{Koen Schakenraad}
\affiliation{Instituut-Lorentz, Leiden University, P.O. Box 9506, 2300 RA Leiden, The Netherlands}
\affiliation{Mathematical Institute, Leiden University, P.O. Box 9512, 2300 RA Leiden, The Netherlands}
\author{Linda Ravazzano}
\affiliation{Instituut-Lorentz, Leiden University, P.O. Box 9506, 2300 RA Leiden, The Netherlands}
\affiliation{Center for Complexity and Biosystems, Department of Physics, University 
of Milan, Via Celoria 16, 20133, Milano, Italy}
\author{Niladri Sarkar}
\affiliation{Instituut-Lorentz, Leiden University, P.O. Box 9506, 2300 RA Leiden, The Netherlands}
\author{Joeri A.J. Wondergem}
\affiliation{Kamerlingh Onnes-Huygens Laboratory, Leiden University, P.O. Box 9504, 2300 RA Leiden, The Netherlands}
\author{Roeland M.H. Merks}
\affiliation{Mathematical Institute, Leiden University, P.O. Box 9512, 2300 RA Leiden, The Netherlands}
\affiliation{Institute of Biology, Leiden University, P.O. Box 9505, 2300 RA Leiden, The Netherlands}
\author{Luca Giomi}
\thanks{Corresponding author: giomi@lorentz.leidenuniv.nl}
\affiliation{Instituut-Lorentz, Leiden University, P.O. Box 9506, 2300 RA Leiden, The Netherlands}

\begin{abstract}
Recent experimental studies have demonstrated that cellular motion can be directed by topographical gradients, such as those resulting from spatial variations in the features of a micropatterned substrate. This phenomenon, known as {\em topotaxis}, is especially prominent among cells persistently crawling within a spatially varying distribution of cell-sized obstacles. In this article we introduce a toy model of topotaxis based on active Brownian particles constrained to move in a lattice of obstacles, with space-dependent lattice spacing. Using numerical simulations and analytical arguments, we demonstrate that topographical gradients introduce a spatial modulation of the particles' persistence, leading to directed motion toward regions of higher persistence. Our results demonstrate that persistent motion alone is sufficient to drive topotaxis and could serve as a starting point for more detailed studies on self-propelled particles and cells.
\end{abstract}

\maketitle

\section{Introduction}
Whether {\em in vitro} or {\em in vivo}, cellular motion is often biased by directional cues from the cell's micro-environment. {\em Chemotaxis}, i.e., the ability of cells to move in response to chemical gradients, is the best known example of this functionality and plays a crucial role in many aspects of biological organization in both prokaryotes and eukaryotes \cite{VanHaastert2004,Swaney2010}. Yet, it has become increasingly evident that, in addition to chemical cues, mechanical cues may also play a fundamental role in dictating how cells explore the surrounding space. {\em Haptotaxis} (i.e., directed motion driven by gradients in the local density of adhesion sites) and {\em durotaxis} (i.e., directed motion driven by gradients in the stiffness of the surrounding extracellular matrix) are well studied examples of taxa driven by mechanical cues \cite{McCarthy1984,Lo2000,Charras2014}. 

{\em In vivo}, cells crawl through topographically intricate environments, such as the extracellular matrix, blood and lymphatic vessels, other cells, etc., that can significantly influence migration strategies \cite{Arcizet2012,Paul2016,Bade2018,Pieuchot2018}. For instance, it has been shown that local anisotropy in the underlying substrate, in the form of adhesive ratchets \cite{Caballero2015,Kim2016,Maout2018} or three-dimensional structures on the subcellular scale \cite{Caballero2015,Heo2017,Park2018}, can lead to directed motion even in the absence of chemical stimuli. More recently, Wondergem and coworkers demonstrated directed migration of single cells using a spatial gradient in the density of cell-sized topographical features \cite{Wondergem2019}. In these experiments, highly motile, persistently migrating cells (i.e., cells performing amoeboid migration) move on a substrate in between microfabricated pillars that act as obstacles and consequently force the cells to move around them. If the obstacles' density smoothly varies across the substrate, the cells have been shown to perform {\em topotaxis}: the topographical gradient serves as a directional cue for the cells to move toward the regions of lower obstacle density. 

Although the precise biophysical or biochemical principles behind topotaxis are presently unknown, its occurrence for cells performing amoeboid migration suggests the possibility of cell-type-indenpendent mechanisms that, separately from the cell's mechanosensing machinery, provide a generic route to the emergence of directed motion. In this article we explore this hypothesis. Using active Brownian particles (ABPs) constrained to move within a lattice of obstacles, we demonstrate that topotaxis can result solely from the spatial modulation of persistence resulting from the interaction between the particles and the obstacles.

ABPs represent a simple stochastic model for self-propelled particles, such as active Janus particles \cite{Ebbens2018}, and cell motility on flat substrates \cite{Selmeczi2005}. ABPs perform persistent self-propelled motion in the direction of the particle orientation in combination with rotational diffusion of this orientation. The motion of active particles has been explored in several complex geometries, including convex \cite{Volpe2011,Fily2014} and nonconvex \cite{Fily2015} confinements, mazes \cite{Khatami2016}, walls of funnels \cite{Galajda2007}, interactions with asymmetric \cite{Angelani2010,Mallory2014} and chiral \cite{Angelani2009} passive objects, porous topographies \cite{Volpe2017} and random obstacle lattices \cite{Reichhardt2014,Zeitz2017,Morin2017,Chepizhko2019}. For a review, see Refs. \cite{Bechinger2016,Zottl2016}. Because of the non-equilibrium nature of active particles, local asymmetries in the environment can be leveraged to create a drift; these particles have been demonstrated to perform chemotaxis \cite{Popescu2018,Saha2019}, durotaxis \cite{Novikova2017}, and phototaxis \cite{Lozano2019}. Furthermore, topographical cues, such as those obtained in the presence of arrays of asymmetric posts \cite{Davies2017,Tong2018} and ratchets consisting of asymmetric potentials \cite{Angelani2011,Ai2013,Ai2014} or asymmetric channels \cite{Ghosh2013,Ao2014,Yariv2014,Katuri2018}, have been shown to produce a directional bias in the motion of active particles reminiscent of those observed for cells.

The paper is organized as follows: in Sec. \ref{sec_model} we present our model for ABPs and their interaction with obstacles. In Sec. \ref{sec_emergence} we show that, in the presence of a gradient in the obstacle density, ABPs drift, on average, in the direction of lower density. The speed of this net drift, here referred to as topotactic velocity, increases as a function of both the density gradient and the persistence length of the ABPs. In Sec. \ref{sec_origin} (numerically) and Sec. \ref{sec_theory} (analytically) we study ABPs in regular obstacle lattices and demonstrate that the origin of topotaxis of active particles can be found in the altered persistence length of the particles in the presence of obstacles.

\section{The model} 
\label{sec_model}
Our model of ABPs consists of disks of radius $R_p$ self-propelling at constant speed $v_{0}$ along the unit vector $\bm{p}=(\cos\theta,\sin\theta)$ and subject to rotational white noise. The dynamics of the particles is governed by the following overdamped equations:

\begin{subequations}
\label{equation_eom}
\begin{align}
\frac{\dd\bm{r}}{\dd t} &=v_{0}\bm{p}+\mu \bm{F}\;, \\
\frac{\dd\theta}{\dd t} &=\sqrt{2D_{r}}\,\xi\;,
\end{align}
\end{subequations}
where $\bm{r}=\bm{r}(t)$ is the position of the particle, $t$ is time, and $\mu $ is a mobility coefficient. The force $\bm{F}=\bm{F}(\bm{r})$ embodies the interactions between the particles and the obstacles. $\xi=\xi(t)$ is a random variable with zero mean, i.e., $\left\langle\xi (t)\right\rangle=0$, and time-correlation $\left\langle\xi  (t)\xi (t^\prime)\right\rangle =\delta (t -t^\prime)$. The extent of rotational diffusion is quantified by the rotational diffusion coefficient $D_{r}$, whereas translational diffusion is neglected under the assumption of large P\'{e}clet number: ${\rm Pe}\gg 1$. Overall, this set-up provides a reasonable toy model for highly motile cells such as those used in experimental studies of topotaxis \cite{Takagi2008,Li2011,Wondergem2019}. For a study on the influence of the P\'{e}clet number on the motion of ABPs around obstacles, see, for example, Ref. \cite{Zeitz2017}.

In free space, (i.e., $\bm{F}=\bm{0}$), ABPs described by Eqs. (\ref{equation_eom}) perform a persistent random walk (PRW) with mean displacement $\left\langle \Delta \bm{r}(t)\right\rangle = 0$ and mean squared displacement: 
\begin{equation}
\label{equation_PRW}
\left\langle \lvert \Delta \bm{r}(t)\rvert^2\right \rangle = 2 v_0^2 \tau_p^2 \left(\frac{t}{\tau_p} +\mathrm{e}^{-t/\tau_p} -1\right)\;,
\end{equation} 
where $\Delta \bm{r}(t) =\bm{r} (t) -\bm{r} (0)$ and $\left\langle \cdots\right\rangle$ represents an average over $\xi$ (see, e.g., Ref. \cite{Zottl2016}). The constant $\tau_{p}=1/D_{r}$, commonly referred to as {\em persistence time}, quantifies the typical timescale over which a particle tends to move along the same direction. Thus, over timescales shorter than the persistence time, $t\ll\tau_p$, ABPs move ballistically with speed $v_0$: $\left\langle \lvert \Delta \bm{r}(t)\rvert^2 \right\rangle \approx (v_0 t)^2$, while over timescales larger than the persistence time, $t\gg\tau_p$, ABPs diffuse, i.e., $\left\langle \lvert \Delta \bm{r}(t)\rvert^2 \right \rangle = 4D t$, with $D =v_0^2 \tau_p/2$ the diffusion coefficient. From $\tau_{p}$, one can define a {\em persistence length}, $l_{p}=v_{0}\tau_{p}$, as the typical distance travelled by a particle before loosing memory of its previous orientation. Consistently, the autocorrelation function of the velocity $\bm{v} =\dd \bm{r}/\dd t$ ($\bm{v}=v_{0}\bm{p}$ in free space) is given by:
\begin{equation}
\label{equation_autocorrelation}
\left\langle \bm{v} (t+\Delta t) \cdot \bm{v} (t)\right\rangle = v_0^2 \mathrm{e}^{-\Delta t/\tau_p}\;.
\end{equation}

Our ABPs roam within a two-dimensional array of circular obstacles of radius $R_{o}$. Following Refs. \cite{Fily2014,Fily2015}, the interactions between particles and obstacles are modeled via a force of the form:
\begin{equation}
\label{equation_wall}
\bm{F}=\begin{cases}-\frac{v_{0}}{\mu} (\bm{p} \cdot \bm{N})\, \bm{N} & \mathrm{if}\: \vert\Delta \bm{r}_{o}\vert\leq R\;, \\
\bm{0} & \mathrm{otherwise}\;, \end{cases}
\end{equation} 
where $\bm{N}$ is a unit vector normal to the obstacle surface, $\vert\Delta \bm{r}_{o}\vert$ is the distance between the obstacle center and the particle center, and the effective obstacle radius $R$ is the sum of the obstacle and the particle radii: $R = R_o + R_p$. Eq. (\ref{equation_wall}) describes a frictionless hard wall force that cancels the velocity component normal to the obstacle surface whenever the particle would penetrate the obstacle, and vanishes otherwise. We stress that the wall force does not influence the intrinsic direction of motion $\bm{p}$. Thus, a particle slides along an obstacle until either the obstacle wall becomes tangential to $\bm{p}$ or rotational diffusion causes the particle to rotate away. This is consistent with experimental observations on self-propelled colloids \cite{Volpe2011} as well as various types of cells \cite{Denissenko2012,Kantsler2013}. For details on the numerical implementation of Eqs. (\ref{equation_eom}) and (\ref{equation_wall}), see Appendix A. In the following Sections, we measure times in units of the the persistence time, i.e., $\tilde{t}=t/\tau_{p}$, and lengths in units of the effective obstacle radius, i.e., $\tilde{\ell}=\ell/R$.  

\section{Results}
\label{sec_results}
\begin{figure}[t]
\centering
\includegraphics[width=\columnwidth]{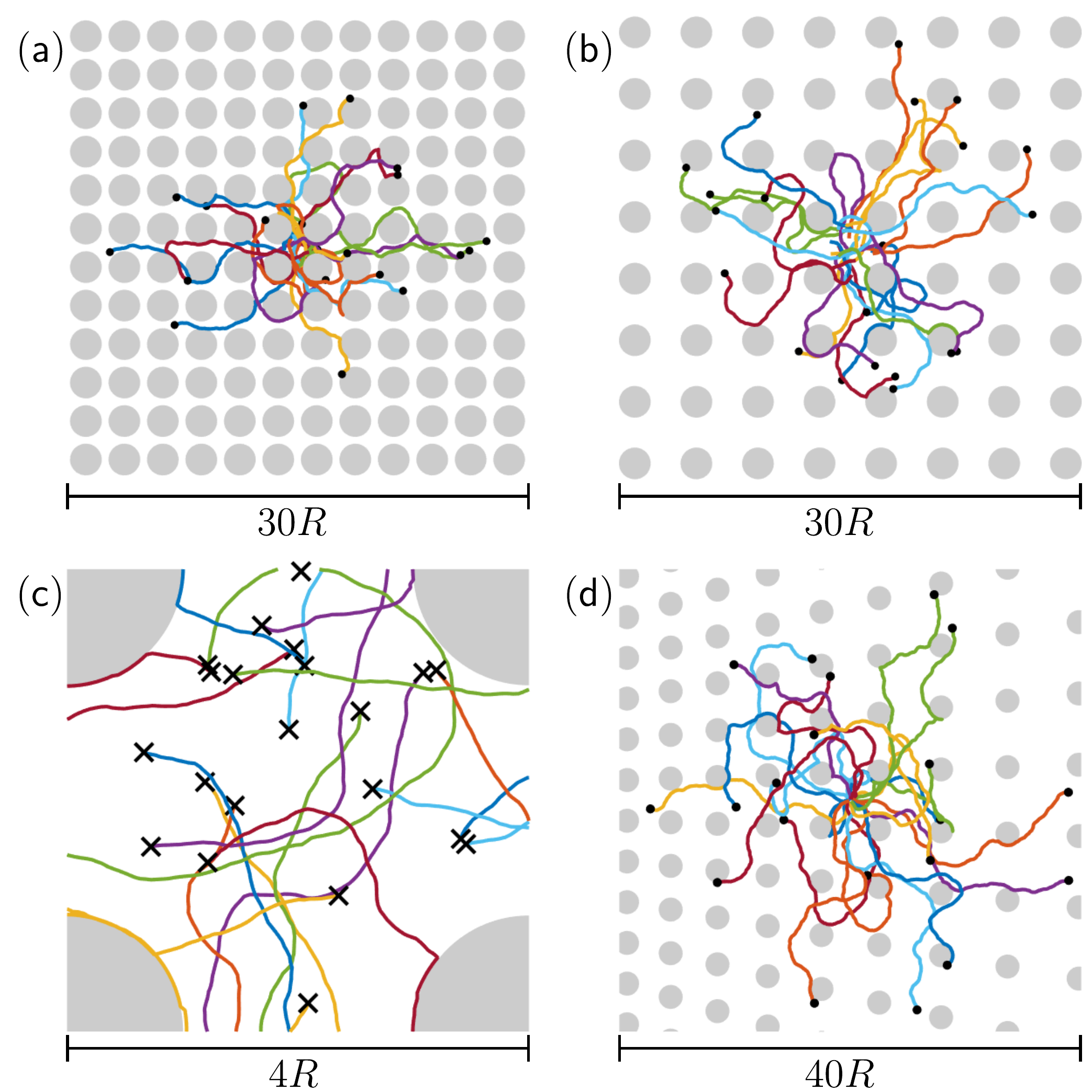}
\caption{\label{fig1} Simulated trajectories of 20 active Brownian particles (ABPs) with persistence length $\tilde{l}_p =v_0 \tau_p /R =5 $ in different lattices of obstacles. The obstacles are graphically represented as disks of radius $R$ and the ABPs as point particles. (a,b) ABPs in regular square lattices of obstacles with center-to-center obstacle spacings $\tilde{d}=d/R  = 2.5$ and $\tilde{d} = 4$ respectively. The particles start at $\tilde{t} = 0$ at a random location in the unit cell of the lattice and are simulated for a total time of $\tilde{t} =3$. (c) The unit cell of the regular square lattice showing the starting points (crosses) of the 20 trajectories in (b). (d) ABPs in a lattice with a linear gradient in obstacle spacing, quantified by a dimensionless parameter $r = 0.07$ (see Appendix B), and $\tilde{d}=5$ at the center of the gradient region. The particles start at $\tilde{t} = 0$ in the origin and are simulated for a total time of $\tilde{t} = 5$.}
\end{figure}

The motion of ABPs in different lattices of obstacles is visualized in Fig. \ref{fig1}. Each panel shows 20 simulated trajectories with persistence length $\tilde{l}_p =5 $. Figs. \ref{fig1}a,b show regular square lattices with dimensionless center-to-center obstacle spacings of $\tilde{d} = 2.5$ and $\tilde{d} = 4$ respectively. In Fig. \ref{fig1}, the obstacles are graphically represented as disks of radius $R$ and the ABPs as point particles. To avoid biasing the statistics of the particle trajectories, ABPs start at a random location inside the unit cell of the regular square lattice (Fig. \ref{fig1}c) at $\tilde{t} =0$ with random orientation. All trajectories are shown for a total time of $\tilde{t} = 3$. Comparing the spreading of the active particles in Fig. \ref{fig1}a with that in Fig. \ref{fig1}b, we observe that the more dense the obstacle lattice is, the more it hinders the diffusion of the active particles. We will quantify this later.

To study topotaxis, we define an irregular square lattice comprising a linear gradient of the obstacle spacing in the positive $x-$direction. The latter is quantified in terms of a dimensionless parameter $r$ representing the rate at which the obstacle spacing decreases as $x$ increases. Thus $r=0$ corresponds to a regular square lattice, whereas large $r$ values corresponds to rapidly decreasing obstacle spacing. Fig. \ref{fig1}d shows this lattice for $r = 0.07$ with 20 particle trajectories, starting in the origin at $\tilde{t} = 0$ with a random orientation, plotted for a simulation time of $\tilde{t} = 5$, where $\tilde{d}=5$ represents the obstacle spacing in the center of the gradient region. The gradient region has a finite width (not visible in Fig. \ref{fig1}d) and is flanked by regular square lattices to the left, with lattice spacing $\tilde{d}_{\mathrm{min}} = 2.1$, and to the right, with lattice spacing $\tilde{d}_{\mathrm{max}} =2\tilde{d}-\tilde{d}_{\mathrm{min}}$. The minimal and maximal obstacle-to-obstacle distances ($\tilde{d}_{\mathrm{min}}$ and $\tilde{d}_{\mathrm{max}}$, respectively) do not depend on the steepness of the gradient, and consequently the width of the gradient region decreases for steeper gradients (larger $r$). For a detailed description of both the regular and gradient lattices as well as an image of the gradient lattice including the regular lattices on the left and right, see Appendix B.

\subsection{The emergence of topotaxis}
\label{sec_emergence}

\begin{figure}[t]
\centering
\includegraphics[width=\columnwidth]{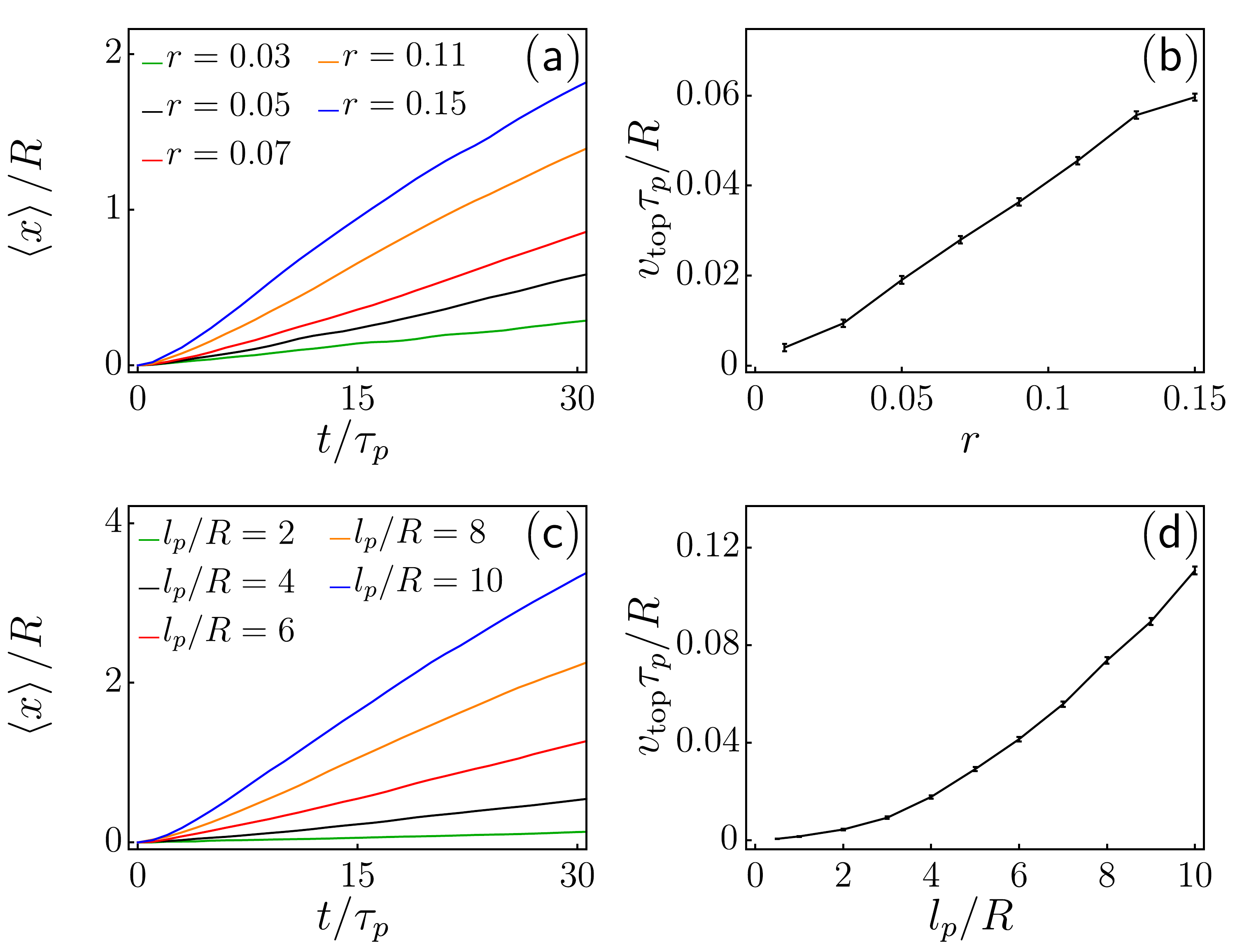}
\caption{\label{fig2} The emergence of topotaxis in density gradient lattices. (a) $\left\langle\tilde{x}\right\rangle = \left\langle x \right\rangle/R$ as a function of time $\tilde{t} = t/\tau_p$ for several values of the density gradient $r$, with $\tilde{d} =d/R =5$ and $\tilde{l}_p =v_0 \tau_p/R =5$. (b) Topotactic velocity, defined in the main text, in the $x$ direction as a function of the density gradient $r$ based on the data in (a). (c) $\left\langle\tilde{x}\right\rangle = \left\langle x \right\rangle/R$ as a function of time $\tilde{t} = t/\tau_p$ for several values of the persistence length $\tilde{l}_p = v_0 \tau_p / R$, with $\tilde{d} =d/R =5$ and $r=0.07$. (d) Topotactic velocity in the $x$ direction as a function of the persistence length $\tilde{l}_p$ based on the data in (c). Data in (a) and (c) represent averages over $10^6$ particles. Error bars in (b) and (d) are given by the standard error of $\left\langle x\right\rangle (t)/(R\; t)$ at $t = 30\;\tau_p$.}
\end{figure}

To quantify topotaxis, we measure the average $x$ and $y$ coordinates, $\left\langle \tilde{x} \right\rangle$ and $\left\langle \tilde{y} \right\rangle$, as a function of time for $10^6$ particles. The results are given for five values of the dimensionless density gradient $r$ in Fig. \ref{fig2}a and Fig. \ref{fig6}a (Appendix C) for $x$ and $y$ respectively. The emergence of topotaxis is clear from Fig. \ref{fig2}a: the active particles move, on average, in the positive $x-$direction, hence in the direction of lower obstacle density. As expected by the symmetry of the lattice, there is no net motion in the $y$ direction independently of the value of $r$ (Fig. \ref{fig6}a, Appendix C). To further quantify topotaxis, we define the topotactic velocity as the average velocity in the positive $x$ direction in a time interval $\Delta t$, $v_{\mathrm{top}} = \left\langle\Delta x\right\rangle/\Delta t$, and evaluate it between $\tilde{t} =0$ and $\tilde{t} = 30$. Figs. \ref{fig2}a and \ref{fig2}b show that $\tilde{v}_{\rm top}$ is approximatively constant in time and proportional to the density gradient $r$. 

Next, we investigate the effect of the intrinsic motion of the ABPs on topotaxis. This intrinsic motion is characterized by the persistence length $l_p =v_0 \tau_p$, which uniquely determines the statistics of the particle trajectory in free space. Namely, if two types of ABPs have different $v_0$ and $\tau_p$, but the same $l_p$, their trajectories have the same statistical properties, even though faster particles move along these trajectories in a shorter time. Fig. \ref{fig2}c shows $\left\langle \tilde{x} \right\rangle$ as a function of time for five values of $\tilde{l}_p$. The speed of topotaxis is again approximately constant in time and increases with $\tilde{l}_{p}$ (Fig. \ref{fig2}d). This trend partly results from the fact that increasing the persistence length corresponds either to an increment in $v_{0}$ or $\tau_{p}$, both resulting into an increase of $\tilde{v}_{0}$. However, Fig. \ref{fig2}d shows that $\tilde{v}_{\mathrm{top}}$ increases faster than linear as a function of $\tilde{l}_p$, suggesting an additional effect caused by the obstacle lattice. As we will see in Sec. \ref{sec_origin}, this effect is caused by the fact that the lattice hinders ABPs with large persistence lengths more than ABPs with smaller persistence lengths. Finally, we note that there is no net motion in the $y$ direction irrespective of the persistence length, as expected by symmetry (Fig. \ref{fig6}b, Appendix C).

\subsection{The physical origin of topotaxis}
\label{sec_origin}

The observed occurrence of topotaxis of ABPs is intuitive, as particles migrate in the direction where there is more available space. However, the mechanism by which ABPs are guided toward the less crowded regions is not obvious from the results in Section \ref{sec_emergence}. To gain more insight into the physical origin of topotaxis, we investigate how particle motility depends on the local obstacle spacing. In doing so, we take inspiration from recent works \cite{Novikova2017,Yu2017,Doering2018} that have shown, in the context of durotaxis, that persistent random walkers, moving in a spatial gradient of a position-dependent persistence length, show an average drift toward the region with larger persistence. As is the case in our system (Sec. \ref{sec_emergence}), this effect is stronger in the presence of larger gradients \cite{Novikova2017,Yu2017}. In order to understand whether or not such a space-dependent persistence might explain the observed topotactic motion, we study and characterize the motion of ABPs in regular square lattices. To do so, we measure the mean squared displacement (MSD) $\left\langle|\Delta \bm{r}(t) |^2\right\rangle$ as a function of time and the velocity autocorrelation function (VACF) $\left\langle\bm{v}(t+\Delta t) \cdot \bm{v}(t)\right\rangle$ as a function of the time interval $\Delta t$ for $10^4$ particles for various lattice spacings $\tilde{d}$ and persistence lengths $\tilde{l}_p$. 

\begin{figure}[t]
\centering
\includegraphics[width=\columnwidth]{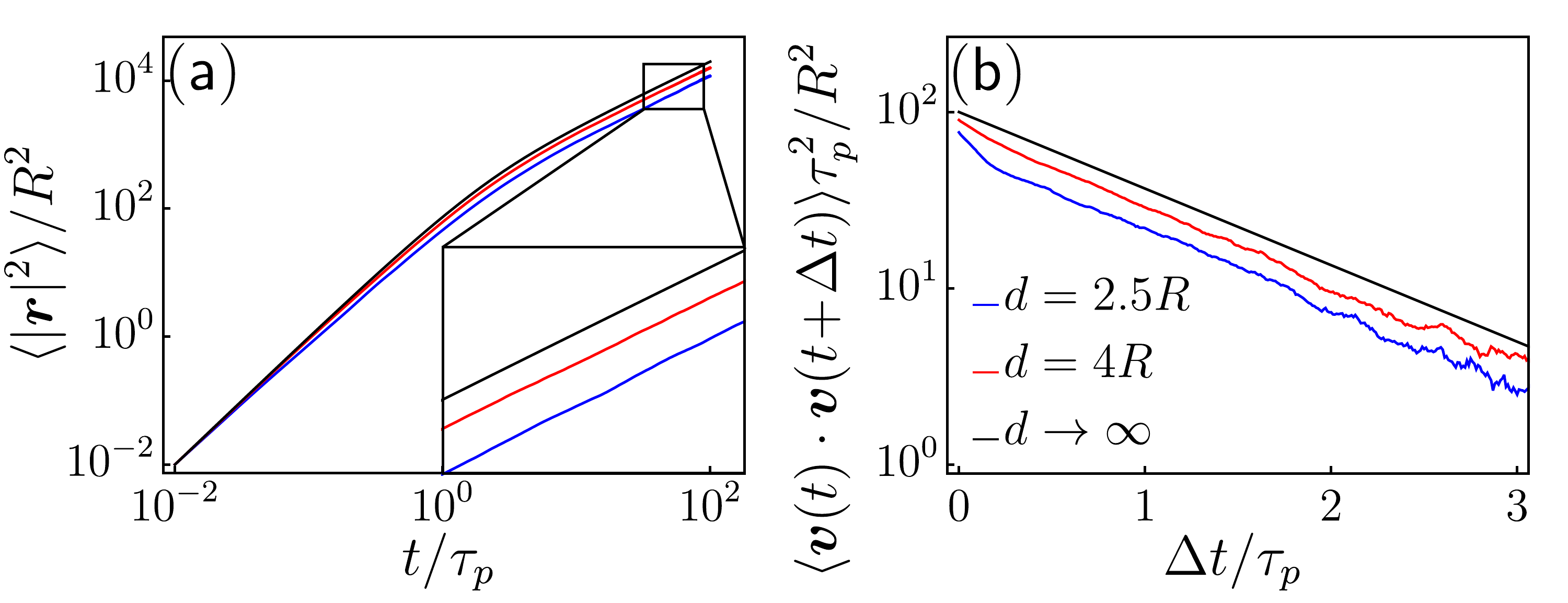}
\caption{Regular square lattices of obstacles modify the effective parameters of the persistent random walk. (a) Dimensionless mean square displacement $\left\langle|\Delta \bm{\tilde{r}} |^2\right\rangle=\left\langle|\Delta \bm{r} |^2\right\rangle/R^2$ as a function of dimensionless time $\tilde{t}=t/\tau_p$ for $\tilde{l}_p =v_0\tau_p/R = 10$ in free space ($d\to\infty$, black line) and in the presence of square lattices with obstacle spacings $\tilde{d} =d/R= 2.5$ (blue) and $\tilde{d} =4$ (red). The effective diffusion coefficient $D_{\mathrm{eff}}$ is obtained from a linear fit to the long-time ($t>\tau_p$) regime of the log-log data. (b) Dimensionless velocity autocorrelation function  $\left\langle\bm{\tilde{v}}(t+\Delta t) \cdot \bm{\tilde{v}}(t)\right\rangle=\left\langle\bm{v}(t+\Delta t) \cdot \bm{v}(t)\right\rangle \tau_p^2/R^2$ as a function of $\Delta \tilde{t}=\Delta t/\tau_p$ for $\tilde{l}_p =v_0\tau_p/R = 10$ in free space ($d\to\infty$) and in the presence of square lattices with obstacle spacings $\tilde{d} = 2.5$ and $\tilde{d} =4$. The effective velocity $v_{\mathrm{eff}}$ and effective persistence time $\tau_{\mathrm{eff}}$ are obtained from a exponential fit to the autocorrelation function. MSD and VACF data represent averages over $10^4$ particles.}
\label{fig3}
\end{figure}

Fig. \ref{fig3}a shows a log-log plot of the MSD for $\tilde{l}_p=10$. The curve with $\tilde{d} \to \infty$ (black) represents the theoretical MSD in free space [Eq. (\ref{equation_PRW})] and exhibits the well-known crossover from the ballistic regime (slope equal to 2) to the diffusive regime (slope equal to 1) around $\tilde{t}=1$ ($t =\tau_p$). The hindrance of the obstacles is evident from the data obtained in regular lattices with $\tilde{d} = 4$ (red curve) and $\tilde{d} = 2.5$ (blue curve), as the MSD is smaller than the MSD in free space at all times (see also the inset). Moreover, the MSD is smaller for the smaller lattice spacing, as we already observed qualitatively in Figs. \ref{fig1}a,b. The hindrance also manifests itself in the short-time ($t < \tau_p$) regime, where the slope of the red and blue curves is slightly smaller than that of the black curve. This indicates that ABPs in obstacle lattices do not move purely balistically, because obstacles prevent them from moving in straight lines.

\begin{figure}[t]
\centering
\includegraphics[width=\columnwidth]{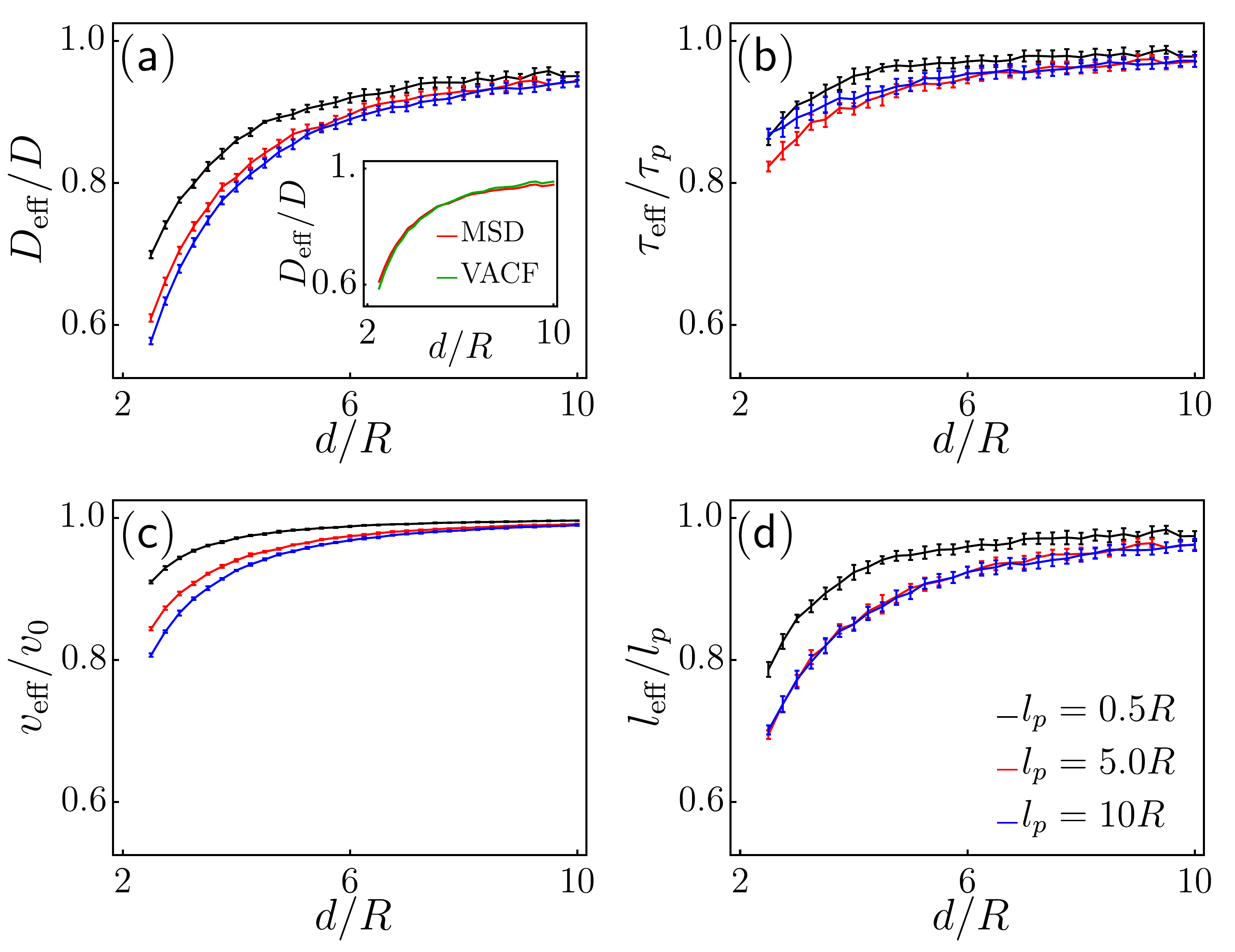}
\caption{Effective parameters of the persistent random walk in regular lattices of obstacles. (a) Normalized effective diffusion coefficient $D_{\mathrm{eff}}/D$, obtained from the mean squared displacement [Eq. (\ref{equation_effective_diffusion})], as a function of the normalized obstacle spacing $\tilde{d} =d/R$ for three values of the normalized persistence length $\tilde{l}_p =v_0 \tau_p /R$. Inset shows $D_{\mathrm{eff}}$ for $\tilde{l}_p =5$ obtained via the mean squared displacement (MSD) and via the velocity autocorrelation function (VACF), using $D_{\mathrm{eff}} = v_{\mathrm{eff}}^2 \tau_{\mathrm{eff}}/2$. (b) Normalized effective persistence time $\tau_{\mathrm{eff}}/\tau_p$, (c) normalized effective velocity $v_{\mathrm{eff}}/v_0$, both obtained from the velocity autocorrelation function [Eq. (\ref{equation_auto_eff})], and (d) normalized effective persistence length $l_{\mathrm{eff}}/l_p = v_{\mathrm{eff}} \tau_{\mathrm{eff}}/(v_0 \tau_p)$ as functions of the normalized obstacle spacing $\tilde{d}=d/R$ for three values of the normalized persistence length $\tilde{l}_p =v_0 \tau_p /R$. Data points represent the average of 10 independent measurements from MSD or VACF data (Fig. \ref{fig3}). The error bars show the corresponding standard deviations.}
\label{fig4}
\end{figure}

The slope of the curves at timescales larger than the persistence time, on the other hand, is independent of the presence of obstacles and equal to 1 (see inset). In other words, even though the motion of the ABPs is hindered by the obstacles at all timescales, the long-time motion remains diffusive, as was also observed for ABPs in random obstacle lattices of low density \cite{Zeitz2017}. Fitting the MSD at long times allows one to define an effective diffusion coefficient $D_{\mathrm{eff}}$, namely:
\begin{equation}
\label{equation_effective_diffusion}
\left\langle|\Delta \bm{r}(t) |^2\right\rangle \xrightarrow [t\gg \tau_p] {}4D_\mathrm{eff}t.
\end{equation}
Fig. \ref{fig3}b shows the velocity autocorrelation function (VACF) on a semilog plot as a function of the time interval $\Delta \tilde{t}=\Delta t/\tau_p$ for $\tilde{l}_p=10$. The $d \to \infty$ curve again represents the theoretical curve in free space and shows exponential decay [Eq. (\ref{equation_autocorrelation})]. Interestingly, in the presence of increasing obstacle densities, hence for smaller lattice spacings $\tilde{d}$, the velocity autocorrelation decreases but remains, to good approximation, exponential. From this numerical evidence, we conclude that the average motion of ABPs in a two-dimensional square lattice can be described as a persistent random walk with an effective velocity $v_{\mathrm{eff}}$ and an effective persistence time $\tau_{\mathrm{eff}}$ \cite{Zeitz2017}, 
\begin{equation}
\label{equation_auto_eff}
\left\langle \bm{v} (t+\Delta t) \cdot \bm{v} (t)\right\rangle  =v_{\mathrm{eff}}^2 \; \mathrm{e}^{-\Delta t/\tau_{\mathrm{eff}}}\;.
\end{equation}
Fig. \ref{fig4} shows $D_{\rm eff}$, $\tau_\mathrm{eff}$ and $v_{\rm eff}$, normalized by their free space values, as a function of the obstacle spacing $\tilde{d}$ for three values of the free space persistence length $\tilde{l}_p$. Starting with the effective diffusion coefficient (Fig. \ref{fig4}a), we observe that, for every value of the persistence length, the effective diffusion coefficient $D_{\mathrm{eff}}$ increases as a function of $\tilde{d}$ until it approaches the free space diffusion coefficient $D$ for large $\tilde{d} $. This is consistent with what we observed in Figs. \ref{fig1}a,b and \ref{fig3}: ABPs on low density lattices spread out more than ABPs on high density lattices. Moreover, the effective diffusion coefficient deviates more from its free space value for large $\tilde{l}_p$ than it does for small $\tilde{l}_p$. This is intuitive because more persistent particles tend to move longer along the same direction and therefore are hindered more in their motion by the obstacle lattice.

The effective persistence time $\tau_{\mathrm{eff}}$ (Fig. \ref{fig4}b) and the effective velocity $v_{\mathrm{eff}}$ (Fig. \ref{fig4}c), both extracted from the velocity autocorrelation function [Eq. (\ref{equation_auto_eff})], show a similar trend: they increase as a function of $\tilde{d}$ until they approach their free space values at high $\tilde{d}$, and they deviate more from their free space values for large $\tilde{l}_p$ than they do for small $\tilde{l}_p$. These data show that the obstacles cause the ABPs, on average, to move slower and turn more quickly. The decreased effective velocity, with respect to free space, is intuitive given the interactions between particles and obstacles [Eq. \eqref{equation_wall}], which slow down the ABPs. The decreased effective persistence time, on the other hand, is less obvious as one could imagine the periodic obstacle lattice to guide ABPs along straight lines. Apparently, this potential guiding mechanism is outcompeted by the fact that encounters of ABPs with individual obstacles at shorter timescales cause them to turn more quickly than in free space. In Sec. \ref{sec_theory} we will study these short-timescale interactions in greater detail.

Combining the effective persistence time (Fig. \ref{fig4}b) and the effective velocity (Fig. \ref{fig4}c) gives the effective persistence length $l_{\mathrm{eff}} = v_{\mathrm{eff}} \tau_{\mathrm{eff}}$ and the effective diffusion coefficient $D_{\mathrm{eff}} = v_{\mathrm{eff}}^2 \tau_{\mathrm{eff}}/2$. The inset of Fig. \ref{fig4}a shows the effective diffusion coefficient for $\tilde{l}_p=5$, calculated both by using the effective persistence time and effective velocity from the velocity autocorrelation function (VACF) and by a direct measurement from the mean squared displacement (MSD). The excellent agreement between $D_\mathrm{eff}$ measured at short and long timescales (using the VACF and MSD respectively) is another indication that the motion of the ABPs in regular square obstacle lattices can indeed be considered to be an effective persistent random walk.

The effective persistence length $l_{\mathrm{eff}} = v_{\mathrm{eff}} \tau_{\mathrm{eff}}$ is plotted in Fig. \ref{fig4}d. As anticipated, the effective persistence length increases with increasing lattice spacing $\tilde{d}$, consistent with findings of ABPs in random obstacle lattices \cite{Zeitz2017} and a model of persistently moving cells in a tissue of stationary cells \cite{Niculescu2015}. This effect increases with the free space persistence length, as the motion of less persistent particles is randomized before they can reach an obstacle. Furthermore, the difference between data with $\tilde{l}_p = 5$ and $\tilde{l}_p = 10 $ is negligible, indicating that the free space persistence length $\tilde{l}_p$ affects particle motion only when it is comparable with the lattice spacing.

These observations, combined with those in Refs. \cite{Novikova2017,Yu2017,Doering2018} which demonstrate a net flux of persistent random walkers toward regions of larger persistence, ultimately explain the origin of topotaxis in our system. ABPs migrate, on average, toward regions of higher persistence, hence to regions of lower obstacle density. Moreover, the dependence of the effective persistence length $l_\mathrm{eff}$ on the free space persistence length $\tilde{l}_p$ (Fig. \ref{fig4}d) justifies the superlinear increase of the topotactic velocity $\tilde{v}_\mathrm{top}$ as a function of $\tilde{l}_p$ (Fig. \ref{fig2}d). In addition to the normal speed-up due to the higher persistence, more persistent particles experience a larger gradient in persistence.

\subsection{Fokker-Planck equation for regular lattices}
\label{sec_theory}
As we explained in Sec. \ref{sec_origin}, topotaxis in our model of ABPs crucially relies on the fact that, even when trapped in an array of obstacles, ABPs still behave as persistent random walkers. The physical origin of this behavior is, however, less clear from the numerical simulations. In this Section, we rationalize this observation using some simple analytical arguments. The probability distribution function $P=P(\bm{r},\theta,t)$ of the position and orientation of an ABP, whose dynamics is governed by Eqs. \eqref{equation_eom}, evolves in time based on the following  Fokker-Planck equation: 
\begin{equation}
{\partial P \over \partial t}= - v_0 \bm{p} \cdot \nabla P - \mu \nabla \cdot (P \bm{F}) + D_R {\partial^2P \over \partial \theta^2}\;, \label{fp}
\end{equation}
subject, at all times, to the normalization constraint:
\begin{equation}
\int \dd\bm{r}\,\dd\theta\,P(\bm{r},\theta,t)=1\;,
\label{prob}
\end{equation}
with $\dd\bm{r}=\dd x\,\dd y$. Eq. \eqref{fp} cannot be solved exactly, but useful insights can be obtained by calculating the rate of change of the mean squared displacement $\left\langle|\Delta \bm{r} |^2\right\rangle$. Here we assume $\bm{r} (0) =\bm{0}$, which yields $ |\Delta \bm{r} |^2 =|\bm{r} |^2 =x^2+y^2$, and
\begin{equation}\label{msd1}
{\partial \left\langle |\bm{r} |^2 \right\rangle \over \partial t}
= \int  \dd\bm{r}\,\dd\theta\, |\bm{r} |^2  {\partial P(\bm{r},\theta,t) \over \partial t}
\end{equation}
Upon substituting Eq. \eqref{fp} in Eq. \eqref{msd1} and integrating by parts, we obtain:
\begin{equation}
{\partial \langle |\bm{r}|^2 \rangle \over \partial t}= 2v_0\langle {\bm r}\cdot {\bm p} \rangle  
+ {2 \mu}\langle {\bm r}\cdot {\bm F} \rangle. \label{msd2} 
\end{equation}
Analogously, the term $\langle \bm{r} \cdot \bm{p} \rangle$ evolves accordingly to: 
\begin{equation}
{\partial \langle {\bm r}\cdot {\bm p} \rangle \over \partial t}= v_0 - D_R\langle {\bm r}\cdot {\bm p} \rangle 
+ \mu\langle {\bm p}\cdot {\bm F} \rangle\;. \label{ret}
\end{equation}
In free space ($\bm{F}=\bm{0}$), Eqs. \eqref{msd2} and \eqref{ret} can be solved exactly to find: 
\begin{equation}
\langle {\bm r}\cdot {\bm p} \rangle_{\bm{F}=\bm{0}} = \frac{v_0}{D_R}\Big(1-\mathrm{e}^{-D_R t}\Big)\;,
\label{re}
\end{equation}
and the mean squared displacement $\langle |\bm{r}|^2\rangle$ given by Eq. \eqref{equation_PRW}. A generic nonzero $\bm{F}$ compromises the closure of the equations, thus making the problem intractable with exact methods. Nevertheless, it is possible to use some simplifying assumptions to obtain intuitive results about $D_{\rm eff}$ and $\tau_{\rm eff}$ (Fig. \ref{fig4}) at short ($t\ll1/D_r=\tau_p$) and long ($t\gg1/D_r=\tau_p$) timescales.

At short timescales, we can assume a particle to be still relatively close to its initial position $\bm{r}(0)=\bm{0}$. Thus one can expand the force in Eq. \eqref{ret} at the linear order in $\bm{r}$, i.e., $\bm{F}(\bm{r})\approx \bm{F}(\bm{0})+\nabla\bm{F}(\bm{0})\cdot\bm{r}$. Evidently, such an expansion is ill-defined for discontinuous forces such as that given by Eq. \eqref{equation_wall}. However, one can imagine to smoothen the force (for instance using a truncated Fourier expansion), without alterning the qualitative picture. By the symmetry of the obstacle lattice, $\bm{F} (\bm{0})=\bm{0}$, $\partial_y F_x(\bm{0}) = \partial_x F_y(\bm{0}) = 0$, and the constant $\partial_x F_x(\bm{0}) = \partial_y F_y(\bm{0}) < 0$, as the horizontal (vertical) component of the force experienced by a particle moving in the positive $x-$direction ($y-$direction), becomes more negative as the particle moves away from the origin. This allows us to write 
\begin{equation}
\left.\langle \bm{p}\cdot\bm{F}\rangle\right|_{t\ll\tau_p} = \frac{\partial F_x}{\partial x}(\bm{0})\,\langle \bm{r} \cdot \bm{p}\rangle\;,
\label{ef}
\end{equation}
and by inserting Eq. \eqref{ef} into Eq. \eqref{ret} we find:
\begin{equation}
\left.{\partial \langle {\bm r}\cdot {\bm p} \rangle \over \partial t}\right|_{t\ll\tau_p}= v_0 - \left(D_r-\mu\frac{\partial F_x}{\partial x}(\bm{0})\right)\langle {\bm r}\cdot {\bm p} \rangle\;. \label{ret2}
\end{equation}
Solving Eq. \eqref{ret2} yields:
\begin{equation}
\langle {\bm r}\cdot {\bm p} \rangle\Big|_{t\ll\tau_p} = \frac{v_0}{D_{r,\mathrm{eff}}}\left(1-e^{-D_{r,\mathrm{eff}} t}\right)\;,
\label{re2}
\end{equation}
with $D_{r,\mathrm{eff}}=D_r-\mu\; \partial_x F_x(\bm{0})>D_r$. By comparing Eq. \eqref{re2} with its free space equivalent [Eq. \eqref{re}], we identify $D_{r,\mathrm{eff}}$ as an increased effective rotational diffusion coefficient. This implies a decreased effective persistence time, consistent with the data in Fig. \ref{fig4}b. The above analysis shows that, to first order, the observed decrease in effective persistence time simply results from the short-time interactions, within a unit cell of the lattice, that cause the particles to turn more quickly. Finally, substituting Eq. \eqref{re2} in Eq. \eqref{msd2} and taking again the first order Taylor expansion for $\bm{F}$ allows one to solve Eq. \eqref{msd2} exactly. Expanding this exact solution at the second order in time yields:
\begin{equation}
\left.\langle |\bm{r}|^2\rangle\right|_{t\ll\tau_p} = v_0^2 t^2\;,
\label{msd3}
\end{equation} 
which is the standard ballistic regime of the mean squared displacement. Hence, the decreased effective velocity observed in Fig. \ref{fig4}c originates from interactions with obstacles at larger timescales ($t\sim\tau_p$, see also Fig. \ref{fig3}b). 

In the long timescale ($t \gg 1/D_r=\tau_p$) the particles reach a diffusive steady state, thus $\partial_{t}\langle {\bm r}\cdot {\bm p}\rangle=0$. Hence, solving Eq. \eqref{ret} for $\langle \bm{r} \cdot \bm{p} \rangle$ and substituting in Eq. \eqref{msd2} yields:
\begin{equation}\label{eq:r2}
\left.{\partial \langle |\bm{r}|^2 \rangle \over \partial t}\right|_{t\gg\tau_p} = \frac{2v_{0}^{2}}{D_{r}}
\left(
1 + \frac{\mu}{v_{0}}\,\langle\bm{p}\cdot\bm{F}\rangle + \frac{\mu D_{r}}{v_{0}^{2}}\,\langle\bm{r}\cdot\bm{F}\rangle
\right)\;,
\end{equation}
As the long time behavior is diffusive, the expression on the right-hand side of Eq. \eqref{eq:r2} is constant and equal to $4D_{\rm eff}$. Now, according to Eq. \eqref{equation_wall}, $\mu({\bm p}\cdot {\bm F}) =\bm{0}$ if $\vert\Delta \bm{r}_{o}\vert>R$, and $\mu ({\bm p}\cdot {\bm F}) =-v_0 (\bm{p}\cdot\bm{N})^2$ otherwise. Thus, $\langle \bm{p} \cdot \bm{F}\rangle <0$. This term shows that diffusion is slowed down because the obstacle force $F$ always slows down the particles (but never accelerates them). Moreover, for more dense obstacle lattices, particles interact with obstacles more often, which explains the observed dependence of $D_\mathrm{eff}$ on the obstacle spacing in Fig. \ref{fig4}a. Analogously, since particles move in an open space and, on average, away from the center, $\langle \bm{r}\cdot\bm{F}\rangle<0$ (i.e., the repulsion forces due to the obstacles are directed more often toward the origin than toward infinity, further slowing down diffusion). Thus $D_{\rm eff} < D$, consistent with our numerical simulations (Figs. \ref{fig3} and \ref{fig4}a).

\section{Discussion and conclusions}
\label{sec_discussion}

In this article we investigated topotaxis, i.e., directed motion driven by topographical gradients, in a toy model of ABPs constrained to move within a two-dimensional array of obstacles of smoothly varying density. We found that ABPs migrate preferentially toward regions of lower density with a velocity that increases with the gradient in the lattice spacing and with the particles' persistence length. In our model, the origin of topotaxis crucially relies on the fact that, even when moving in a lattice of obstacles, ABPs still behave as persistent random walkers, but with renormalized transport coefficients: $\tau_{\rm eff}$ and $v_{\rm eff}$. As these depend on the topography of the substrate, here quantified in terms of lattice spacing, topographical gradients result into spatially varying persistence in the motion of the particles, which in turn drives directed motion toward regions of larger persistence \cite{Novikova2017,Yu2017,Doering2018}. We note that the motion we report here, just like the durotactic motion described in Refs. \cite{Novikova2017,Yu2017}, is perhaps better described as a ``kinesis'' than as a ``taxis'', because the underlying mechanism of transport is a nondirectional change in behavior induced by a purely positional cue. This in contrast to the true directional bias underlying, for instance, chemotaxis of \emph{E. coli} \cite{Berg1972} which leads to significantly more efficient transport \cite{Doering2018}.

Several questions remain open to future investigation. For instance, how is the picture affected by translational diffusion? Is topotaxis robust against competing directional cues, such as chemotaxis \cite{Wondergem2019}? How sensitive is the performance of topotaxis with respect to the obstacles' shape \cite{Volpe2011,Davies2017,Tong2018}, the type of motion (e.g., persistent random walk, run-and-tumble, L\'evy walk, etc. \cite{Volpe2011,Berdakin2013,Mijalkov2013,Khatami2016,Volpe2017,Bertrand2018}), and the details of particle-obstacle interactions \cite{Chepizhko2013,Takagi2014,Morin2017,Davies2017,Jakuszeit2019}? 
Another interesting setting of the problem could be obtained by considering random arrangements of obstacles, where, unlike in the lattices studied here, particles can be trapped into convex-shaped features that can significantly alter their motion \cite{Zeitz2017,Volpe2017}. 

Finally, although here we demonstrated that topotaxis can be solely driven by the interplay between topographical gradients and persistent random motion, whether this is sufficient to explain cellular topotaxis remains an open problem. A quantitative comparison between our numerical data and experiments on highly motile cells \cite{Wondergem2019} shows, in fact, discrepancies that could be ascribed to the enormously more complex interactions between cells and their environment. Specifically, the topotactic velocity in our simulations is in the order of $1 \%$ of the intrinsic particle speed (Fig. \ref{fig2}), whereas in the experiments on cells this ratio is approximately $5\%$, provided that the obstacles are not spaced further apart than the cell size \cite{Wondergem2019}. In order to better understand this surprising efficiency, the topotactic response of several types of persistently moving cells, such as amoeba \cite{Bretschneider2016}, cancer cells \cite{Sahai2007}, or leukocytes \cite{Friedl2001}, could be compared. On the theoretical side, we are currently addressing the problem using more biologically-realistic models of cell motility based on the cellular Potts model \cite{Maree2012,Niculescu2015}, which allow explicitly taking into account effects such as the resistance of cells against deformations or adhesion between cells and obstacles. 

\acknowledgments
This work was supported by funds from the Netherlands Organization for Scientific Research (NWO/OCW), as part of the Frontiers of Nanoscience program (L.G.), the Netherlands Organization for Scientific Research (NWO-ENW) within the Innovational Research Incentives Scheme  (R.M.H.M.; Vici 2017, No. 865.17.004), and the Leiden/Huygens fellowship (K.S.).

\appendix

\section{Numerical methods}
We numerically generate particle trajectories that perform a persistent random walk by discretizing the equations of motion as follows \cite{Novikova2017}: a particle starts at position $\bm{r}_0$ at $t = 0$, after which the particle is moved by a distance $v_{0}\Delta t$ in a random initial direction $-\pi<\theta_1<\pi$, such that the new position is $\bm{r}_1 = \bm{r}_0 + v_0\Delta t \; \bm{p} (\theta_1)$. For all subsequent time steps, the angle at time step $n$, $\theta_n$, is updated by adding a small deviation angle to the angle of the previous time step, $\theta_n = \theta_{n -1} +\delta\theta$. Here, $-\pi<\delta\theta<\pi$ is
extracted randomly from a Gaussian distribution with mean $0$ and variance $\sigma^{2}=2\Delta t/\tau_{p}$ using the Box-Muller transform. The new position of the particle, $\bm{r}_n$, is then found by $\bm{r}_n = \bm{r}_{n -1} +  v_{0}\Delta t \; \bm{p} (\theta_n)$, with $\bm{r}_{n -1}$ the position at the previous time step.

If the update step moves the particle into an obstacle, however, the particle-obstacle force [Eq. (\ref{equation_wall})] is triggered. In that case, the normal component of the attempted displacement is subtracted, and the actual displacement is given by the tangential component of the attempted displacement, $\bm{r}_n =\bm{r}_{n -1} + v_{0}\Delta t  \; \Big(\bm{p}(\theta_n) \cdot \bm{T}\Big) \; \bm{T}$, with $\bm{T}$ the tangent unit vector of the obstacle surface at the point of the surface closest to $\bm{r}_{n -1}$. We choose the time step $\Delta t$ such that it is much smaller than the persistence time, $\Delta t\ll\tau_p$, and such that every displacement is much smaller than the obstacle radius, $v_0\Delta t\ll R$. In all reported simulations we have used $\Delta t = 0.01\tau_p$.

\section{Obstacle lattices}
\begin{figure}[t!]
\includegraphics[width=\columnwidth]{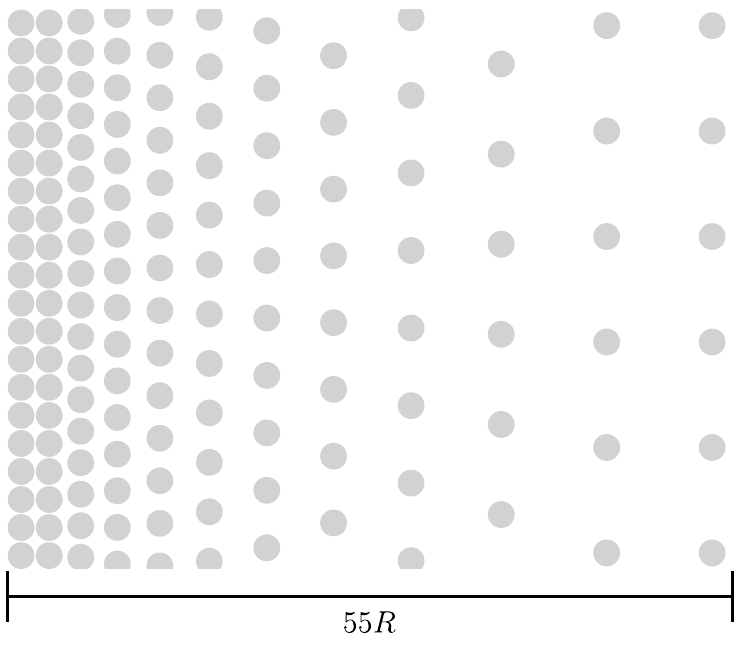}
\caption{Snapshot of the gradient lattice as described in Appendix B. The gradient region is characterized by $r = 0.15$ and $\tilde{d}=d/R = 5$. The obstacles are graphically represented as disks of radius $R$. The lattice spacing varies from $\tilde{d}_\mathrm{min}=2.1$ to $\tilde{d}_\mathrm{max}=7.9$ over the $x$ range $[x_\mathrm{min},x_\mathrm{max}] = [-19,19]$. The gradient region is flanked by a regular square lattice with $\tilde{d}=\tilde{d}_\mathrm{min}$ on the left ($x<x_\mathrm{min}$) and by a regular square lattice with $\tilde{d}=\tilde{d}_\mathrm{max}$ on the right ($x>x_\mathrm{max}$). Only the first two columns of both (infinitely large) regular lattices are shown.}
\label{fig5}
\end{figure}
We define a regular square lattice of obstacles with the coordinates of the centers of the obstacles given by
\begin{subequations}
\label{equation_regular}
\begin{align}
x(n,m) & = nd+\frac{d}{2} \\
y(n,m) &= md+\frac{d}{2}
\end{align}
\end{subequations}
where $n,m\in \mathbb{Z}$ are the obstacle numbers and $d$ is the distance between the centers of two neighboring obstacles. The term $d/2$ is added to make sure that the origin of the coordinate system is in the middle of four obstacles. An illustration of this lattice is given in Figs. \ref{fig1}a,b.

We define an irregular square lattice with a linear gradient of the obstacle spacing in the positive $x$ direction. The gradient region has a finite width, is centered in the origin, and is flanked by regular square lattices to the left and to the right. The coordinates of the centers of the obstacles in the gradient region are given by
\begin{subequations} 
\label{equation_gradient}
\begin{align}
x(n,m) & = \frac{d}{1-e^{-r}}(e^{rn}-1)+\frac{d}{2} \\
y(n,m) & = d(m+\frac{1}{2})e^{rn}\label{equation_gradient_b}
\end{align}
\end{subequations}
where $n,m\in \mathbb{Z}$ are again the obstacle numbers, $d$ is the distance between the centers of obstacles with $(n,m) = (0,0)$ and $(n,m) = (-1,0)$ (i.e., the lattice spacing in the origin), and $r$ is a dimensionless number that quantifies the gradient in the obstacle spacing.

Eq. (\ref{equation_gradient}) represents an obstacle lattice where the lattice spacing depends exponentially on the horizontal obstacle number $n$, such that $x (n,m) -x (n -1,m) =d e^{rn}$ and $y (n,m) -y (n,m -1) =d e^{rn}$. This exponential gradient in the obstacle spacing, as a function of the obstacle number $n$, leads to a linear gradient in the obstacle spacing as a function of the horizontal coordinate $x$. This can be seen by calculating the difference in obstacle distance between two adjacent pairs of obstacles, divided by the distance between those two pairs,
\begin{equation}
\label{equation_gradient2}
\frac{\Big(x (n +1) -x (n)\Big) -\Big(x (n) -x (n -1)\Big)}{x (n) -x (n -1)} = e^{r} -1,
\end{equation}
which is independent of $n$, as required for a linear gradient. In the limit of $r \to 0$, Eqs. (\ref{equation_gradient}) reduce to the regular square lattice given in Eqs. (\ref{equation_regular}).

The gradient lattice is cut off on the left side at $x_{\mathrm{min}} <0$, where the vertical distance between two neighboring obstacles, $y(n,m) - y (n,m-1) $, would otherwise become smaller than a minimal distance $d_{\mathrm{min}} =2.1 R$. At the first column of obstacles for which this is the case, the vertical coordinates [Eq. \eqref{equation_gradient_b}] are replaced by $y (n,m) =md_\mathrm{min} +d_\mathrm{min}/2$. To the left of this transition column ($x<x_{\mathrm{min}}$), a regular obstacle lattice with spacing $d_{\mathrm{min}}$ is placed such that the transition column is part of this regular lattice.

On the right side the gradient lattice is cut off at $x_{\mathrm{max}} = -x_{\mathrm{min}}$. To the right of this cut-off ($x>x_{\mathrm{max}}$), a regular obstacle lattice with spacing $d_{\mathrm{max}} =2d-d_{\mathrm{min}}$ is placed such that the horizontal distance between the rightmost column of the gradient lattice and the leftmost column of the regular lattice is equal to $d_\mathrm{max}$. Thus, the gradient lattice connects two regular square lattices of lattice spacings $d_{\mathrm{min}}$ and $d_{\mathrm{max}}$. The width of the gradient region, $2 x_{\mathrm{max}}$, then depends on the gradient parameter $r$. For an illustration of the gradient lattice for $r = 0.15$ and $\tilde{d} = d/R = 5$, see Fig. \ref{fig5}.

\begin{figure}[t!]
\includegraphics[width=\columnwidth]{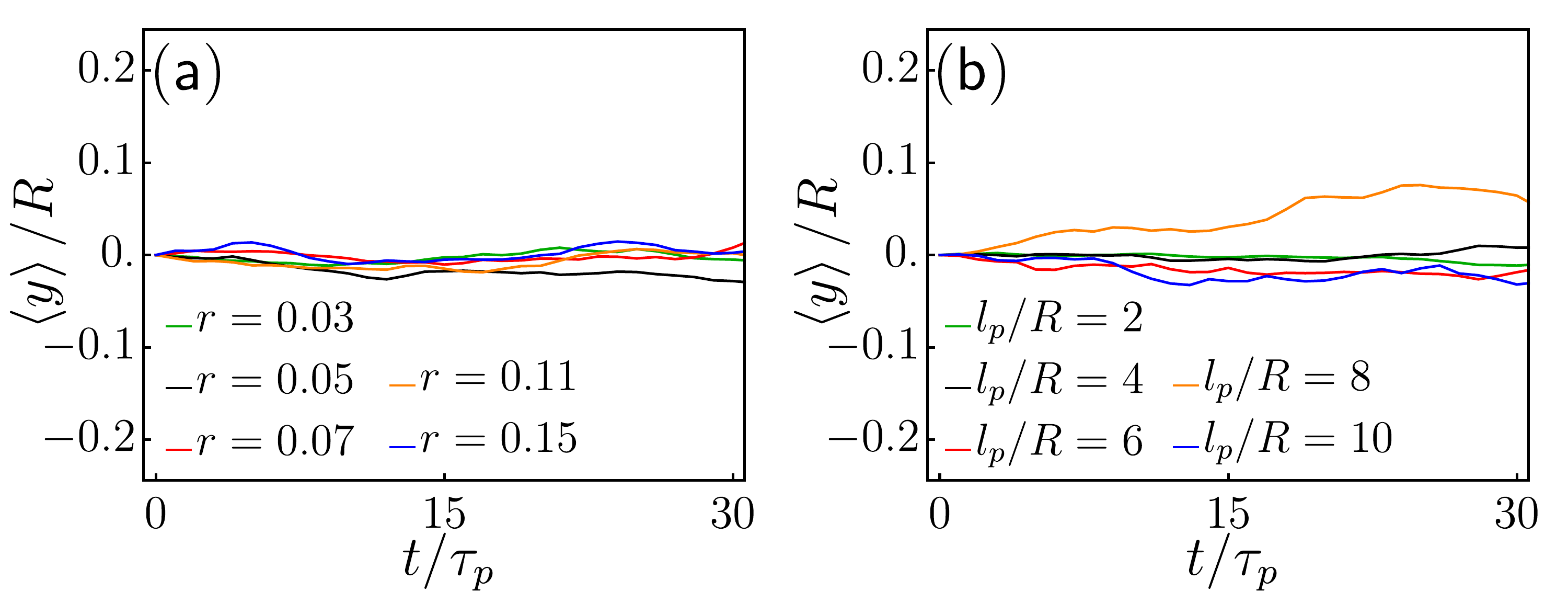}
\caption{There is no average drift in the $y $ direction in density gradient lattices. (a) $\left\langle\tilde{y}\right\rangle = \left\langle y \right\rangle/R$ as a function of time $\tilde{t} = t/\tau_p$ for five values of the density gradient $r$, with $\tilde{d} =d/R =5$ and $\tilde{l}_p =v_0 \tau_p/R =5$. (b) $\left\langle\tilde{y}\right\rangle = \left\langle y \right\rangle/R$ as a function of time $\tilde{t} = t/\tau_p$ for five values of the persistence length $\tilde{l}_p = v_0 \tau_p / R$, with $\tilde{d} =d/R =5$ and $r =0.07$. }
\label{fig6}
\end{figure}

\vspace{-8pt}
\section{Average motion in $y$}
We plot $\left\langle\tilde{y}\right\rangle (\tilde{t})$ of $10^6$ particles moving in a density gradient lattice with $\tilde{d} =5$, starting in the origin with a random orientation, for several values of the dimensionless density gradient $r$ in Fig. \ref{fig6}a, and for several values of the persistence length $\tilde{l}_p$ in Fig. \ref{fig6}b. As expected, there is no average drift in the $y$ direction. The fluctuations in $\left\langle y\right\rangle$ are in the order of $10\%$ of the effective radius $R$.

\end{document}